\documentclass[%
 reprint,
 amsmath,amssymb,
 aip,
 cha,    
]{revtex4-2}

\usepackage{graphicx}
\usepackage{dcolumn}
\usepackage{bm}

\usepackage{enumitem}
\usepackage{color}
\usepackage{xcolor}
\usepackage{soul}

\usepackage{array}
\usepackage{geometry}
\usepackage{booktabs}
\usepackage{xcolor}
\usepackage{colortbl}

\begin{document}

\title{Bio-inspired AI: \\Integrating Biological Complexity into Artificial Intelligence
}

\author{Nima Dehghani}
 \email{nima.dehghani@mit.edu}
\affiliation{McGovern Institute for Brain Research, MIT, Cambridge, MA 02139, USA.}%
\affiliation{Allen Discovery Center at Tufts University, 200 Boston Ave., Suite 4600, Medford, MA 02155, USA}
\affiliation{Center for Brains, Minds, and Machines,  MIT, Cambridge, MA 02139, USA.}
\altaffiliation[Current Address: ]{McGovern Institute for Brain Research, MIT, Cambridge, MA 02139, USA.}

\author{Michael Levin}
\affiliation{Allen Discovery Center at Tufts University, 200 Boston Ave., Suite 4600, Medford, MA 02155, USA}%
\affiliation{Wyss Institute for Biologically Inspired Engineering, Harvard University, Boston, MA 02115, USA}

\date{\today}

\begin{abstract}
The pursuit of creating artificial intelligence (AI) mirrors our longstanding fascination with understanding our own intelligence. From the myths of Talos to Aristotle's logic and Heron's inventions, we have sought to replicate the marvels of the mind. While recent advances in AI hold promise, singular approaches often fall short in capturing the essence of intelligence. This paper explores how fundamental principles from biological computation—particularly context-dependent, hierarchical information processing, trial-and-error heuristics, and multi-scale organization—can guide the design of truly intelligent systems. By examining the nuanced mechanisms of biological intelligence, such as top-down causality and adaptive interaction with the environment, we aim to illuminate potential limitations in artificial constructs. Our goal is to provide a framework inspired by biological systems for designing more adaptable and robust artificial intelligent systems.
\end{abstract}


\maketitle

\section{Historical Context}
The pursuit of artificial intelligence has deep historical roots. Philosophers like Thomas Hobbes, with his mechanical theory of thought, and visionaries such as Blaise Pascal and Gottfried Leibniz ignited the dream of intelligent machines, foreshadowing the notion that intelligence could arise from complex, rule-bound manipulations.

Subsequent centuries saw renewed efforts to formalize intelligence. Charles Babbage's programmable machines and Ada Lovelace's realization that machines could manipulate symbols—not just numbers—laid the groundwork for modern computation. The 20th century, from formal logic to Alan Turing's foundational work, witnessed the birth of Artificial Intelligence. Despite setbacks, this era produced remarkable successes in specialized tasks; machines achieved impressive feats in chess (IBM DeepBlue \cite{Ccampbell2007}) and Go (AlphaGo \cite{Silver2016}), employing powerful search algorithms that differed significantly from the adaptable, context-sensitive intelligence of biological systems.

Recent advances in neuroscience and computing power have ignited excitement for neuro-inspired computer vision \cite{Cox2014}. However, the elusive goal of Artificial General Intelligence (AGI) remains, as the definition of intelligence evolves and benchmarks like the Turing test shift our expectations \cite{Mitchell2024a,Mitchell2024b}. Promising strategies like symbolic computation \cite{Newell1961} ultimately fell short, possibly due to the lack of a vast knowledge base reflecting real-world understanding \cite{McCarthy1987}. Neural networks, once met with skepticism, now flourish with increased computational power, yet even the most sophisticated networks still fall short of the adaptive, context-dependent intelligence observed in biological systems.

These limitations have prompted researchers to explore alternative paradigms inspired by the adaptability and complexity of biological systems, which we will discuss in the following section.

\section{Bio-inspired AI: Current State and Limitations}
Biological intelligence, refined through evolutionary processes, offers an alternative paradigm emphasizing adaptability and context sensitivity. Evolution acts as a natural tinkerer \cite{Jacob1977}, incrementally building complex systems that flexibly respond to environmental challenges, leading to hierarchically structured architectures that prioritize context-dependent information processing.

Biological intelligence is not confined to neural systems. Research shows that even single-celled organisms exhibit information processing and adaptive behavior. For instance, \textit{Paramecium}, a eukaryotic unicellular ciliate, can integrate sensory inputs and modify its behavior accordingly \cite{Brette2021}. Plants also demonstrate sophisticated behaviors like signal transduction and environmental responsiveness without a nervous system, challenging traditional notions of cognition \cite{Trewavas2005,Colaco2022}.

Moreover, life has been navigating various problem spaces—metabolic, transcriptional, physiological, and anatomical—long before nerves and muscles appeared \cite{Fields2020,Fields2022}. This indicates that intelligence predates multicellularity \cite{Lyon2015,Baluska2019,Baluska2022} and is rooted in fundamental biological processes. Cells' problem-solving abilities in physiological and metabolic domains enable cell collectives to navigate morphospace—an abstract multidimensional space of possible biological forms and functions—by relying on internal molecular pathways and gene-regulatory networks for learning and decision-making \cite{Biswas2021,Biswas2023,Timsit2021,Gunawardena2022,Gershman2021}.

Thus, a path toward artificial general intelligence involves not merely replicating neural architectures but implementing the multi-scale, hierarchical organization found in life. In such systems, every level—from molecules to cells to organisms—possesses competencies and engages in adaptive information processing, a concept known as ``polycomputing'' \cite{Bongard2023,Levin2024}. By embracing this broader view of intelligence, we can explore bio-inspired designs that harness the inherent adaptability and context sensitivity of biological systems.

Recent years have seen a shift towards bio-inspired designs in developing intelligent machines. Projects like RoboBee \cite{Ma2013} and RoboRay \cite{Park2016}, which mimic insect flight and combine biological with electronic elements, underscore the importance of understanding material constraints for creating flexible, lightweight systems. Effective communication among components and processing power are key to their success. More directly addressing intelligence, neuromorphic computing aims to achieve brain-like energy efficiency with spiking neural network designs \cite{Merolla2014,Furber2016}. In algorithms, deep learning \cite{LeCun2015} and recurrent networks \cite{Hopfield1982} loosely mimic aspects of biological network structure.


However, much of intelligent problem-solving in biology occurs without spiking activity or neurons \cite{Lyon2006,Ball2016}, challenging the AI focus on neural networks. Intelligence can emerge from different biological substrates and mechanisms. For instance, slime molds like Physarum polycephalum can find the shortest path through a maze to reach food sources, effectively solving complex optimization problems without any neural structures \cite{Tero2010}. Plants exhibit sophisticated behaviors such as opening and closing stomata in response to environmental conditions and adjusting growth patterns toward light sources through hormonal signaling pathways \cite{Trewavas2005}. Social insects like ants and termites coordinate colony-level activities such as foraging, nest building, and defense through pheromone communication and emergent behaviors, without relying on neuronal processing akin to spiking neurons \cite{Camazine2001}. Even single-celled organisms like bacteria demonstrate chemotaxis and quorum sensing, adjusting their actions based on chemical gradients and population density, reaching optimal census \cite{Bassler2002,Taillefumier2015}. Remarkably, genetically modified bacteria can act as cell-based biocomputers, solving mathematical problems like identifying prime numbers \cite{Bonnerjee2024}.

These examples illustrate that intelligence in nature often arises from hierarchical, context-dependent interactions within and between living subsystems at multiple scales—not solely from neural computations. They challenge the assumption that replicating neuronal structures is the only path toward artificial intelligence and highlight the potential of alternative biological mechanisms.

Despite some bio-inspired advances, understanding the computational limits imposed by biological organization remains elusive. Current AI approaches, focused on mimicking ``behavioral function'', often overlook either complex hierarchical architectures or adaptive environmental interactions. This stems from a historical emphasis on functional aspects of intelligence \cite{Rosenblueth1943}, where the functional output is paramount but intrinsic organization and external context are secondary. While achieving functionality is desirable, replicating raw computational power incurs high energy costs if we do not understand  intrinsic organization and external context.

Early attempts to incorporate goal-directed (teleological) behavior via feedback loops offered a path toward adaptation \cite{Rosenblueth1943,Simon1962,Simon1969}. W. Grey Walter's \textit{Machina Speculatrix} and \textit{Machina Docilis} tortoises \cite{Walter1950} could navigate environments using simple sensors, displaying purposeful behaviors. However, purely behavioral approaches failed to advance AI beyond narrow tasks. The missing piece may be the hierarchical organization and multi-scale interactions found in biological systems, enabling context-dependent functional algorithms.

In biology, context dependency involves top-down information exchange unique to multi-scale organization \cite{Walker2012}, where higher levels modulate lower-level components \cite{Noble2012}, allowing dynamic adaptation. While information is physical and subject to thermodynamic limits \cite{Landauer1991,Landauer1961,Berut2012,Parrondo2015,Toyabe2010}, biological systems exhibit feedback where macroscopic information influences microscopic dynamics \cite{Walker2012}. This hints at strong causal emergence \cite{Hoel2013}, where higher-level properties exert causal powers not reducible to lower-level interactions \cite{Noble2012}.

Furthermore, biological systems often perform polycomputing—the ability of the same substrate to perform multiple computations simultaneously \cite{Bongard2023,Levin2024}. Unlike traditional computers, living systems components carry out diverse functions concurrently; for example, a protein may participate in metabolism, signal transduction, and structural support. This multifunctionality is a hallmark of biological efficiency and adaptability.

This interplay of information and energy allows goal-directed, context-sensitive processing to reconfigure microscopic dynamics. Future quantiative biological models must translate macroscopic goals into computational rules at network, unit (cells), and intracellular scales \cite{Dehghani2024}. Conventional AI often seeks intelligence emergence from components lacking world models or goal-directed behavior. In contrast, biology exhibits problem solving at every level, with behavior resulting from cooperation and mutual shaping of diverse modules across scales. Every component has homeostatic goal-directedness, and higher levels can modulate the energy landscape of sub-levels to harness autonomy toward larger goals.

Biological systems also operate within unreliable mediums due to unavoidable noise, forcing them to reinterpret information on-the-fly, prioritizing saliency and real-time adaptation over data fidelity \cite{Levin2024}. This frames information in biology as fundamentally semantic, carrying meaning and purpose, not merely randomness as in Shannon's information theory \cite{Shannon1948,Smith2000}. Thus,  it reflects information's physical nature, offering a framework distinct from the vague ``brain (or cell) is a computer'' analogy \cite{Dehghani2024}. Biological systems blur the separation of machine and data, as components can modify their structure and function based on information processing. 

While bio-inspired AI holds promise, we must be mindful of its limitations. The complexity of biological systems poses challenges in replicating multiple organizational levels. Strictly adhering to biological models risks overfitting; biology should inspire, not constrain, AI design. Scaling bio-inspired systems to real-world applications involves engineering inherently complex context-dependent interactions.

\section{Conceptual Foundations for Bio-inspired AI}
\subsection{Contextual Information Processing in Biology}
Biological intelligence relies on context, non-locality, and adaptive feedback. Unlike deterministic algorithms seeking optimal solutions through exhaustive search or predefined rules, biological systems interpret information based on current context. Meaningful information emerges from interactions across multiple scales within a hierarchical architecture, without a direct one-to-one correspondence between microscopic dynamics (e.g., individual neuron activity) and information flow.

This framework mirrors gene expression in cells, where DNA functions not as a rigid blueprint but as a dynamic backbone influenced by the cellular environment \cite{Ball2016, Walker2012}. Genes interact and respond to external signals, leading to the expression or suppression of certain genes; these interactions, rather than gene products alone, carry information and realize possibilities based on genetic composition \cite{Smith2000}.

The key insight is that biological systems do not operate on fixed instructions but adaptively process information based on context. Evolution acts as a natural tinkerer \cite{Jacob1977}, incrementally building complex systems that flexibly respond to environmental challenges. This principle has inspired fields like evolutionary robotics, applying evolutionary concepts to robot design and leading to intelligent behaviors differing from those produced by traditional algorithms \cite{Pfeifer2007}. Similarly, within organisms, processes like morphogenesis enable the development of complex structures through context-dependent cellular interactions, while behavioral circuits allow adaptive responses during an organism's lifetime.

\subsection{Trial and Error as a Fundamental Strategy}
In complex environments, organisms cannot rely on precomputed optimal solutions; instead, they employ heuristics derived from experience, particularly trial-and-error methods, to adapt to new challenges \cite{Pearl1984, Radnitzky1989}. This approach enables effective exploration without exhaustive search, favoring adaptability over perfection. Ashby demonstrated that parallel trial-and-error strategies can rapidly solve high-dimensional problems infeasible for serial exhaustive methods \cite{Ashby1960}. Similarly, chess masters use experience-based heuristics to focus on promising moves, navigating vast possibilities without calculating every option \cite{Simon1962a}. These non-optimal yet effective strategies are characteristic of biological intelligence.




\subsection{Maintaining Stability Through Hierarchical Organization}
A critical question arises: How does a system exhibiting context-dependent, non-local behavior remain stable? Ashby's Law of Requisite Variety provides an answer \cite{Ashby1958}: to effectively control a system or adapt to an environment, the internal control mechanism must possess a variety at least equal to that of the system-environment complex. 

However, simply increasing components and connections can lead to instability; Gardner and Ashby showed that excessive connectivity causes chaotic dynamics \cite{Gardner1970}. Biological systems address this challenge through modular organization—subsystems performing specific functions—maintaining stability while allowing complex behaviors \cite{May1972, Dehghani2024}. For example, modules interact but their interactions become fixed over time to ensure consistent functionality \cite{Cohen1985}. This modular and hierarchical organization enables robustness and flexibility, allowing systems to adapt without becoming unstable. Moreover, competitive dynamics can shape stable collective information processing, with modules recruited as needed \cite{Daniels2016}.



\subsection{Hierarchy and Information Abstraction}
Evolution has embraced hierarchical organization to build adaptive systems \cite{Dehghani2024}. Simple organisms exhibit rudimentary behaviors, while more complex ones display intricate behaviors enabled by deeper hierarchical structures like nervous systems. Comparative neuroanatomy across species reveals increasing structural hierarchy alongside more sophisticated behaviors.

Hierarchy involves organizing nested component for information abstraction and efficient processing \cite{dehghani2024physical}, akin to how deep learning networks use multiple layers to extract higher-level features \cite{Lin2017}. Hierarchical structures facilitate renormalization processes, summarizing detailed lower-level information at higher levels to handle complexity effectively.

Importantly, hierarchical organization provides modularity, aiding in repair and improvement without disrupting the entire system. Simon's parable of the two watchmakers illustrates this point \cite{Simon1962, Simon1969}: while Tempus builds watches serially (in a non-modular fashion) and must restart upon interruption, Hora uses modules, making the process more efficient and resilient.

Evolutionary tinkering crafts a modular structure that increases the system's requisite variety as its functional repertoire grows. This principle extends to biological systems and can inform robust AI design. By adopting a modular, hierarchical approach, AI systems can achieve greater adaptability and resilience.

\subsection{Beyond Deterministic Algorithms}
This understanding challenges attempts to formalize intelligence purely through induction or deterministic algorithms. Bayesian inference, relying on hierarchical generative models and prior assumptions \cite{Tenenbaum2011}, cannot achieve the contextual, non-optimal generality of biological intelligence \cite{Deutsch2012} due to its reliance on predefined priors, which limits exploration and adaptability. Trial and error, likely the most fundamental knowledge-gathering strategy, underpins even scientific discovery \cite{Popper1959, Deutsch2013}. Complex intelligent behavior arises through continuous interaction with and modification of the environment. Biological systems exemplify this through constant feedback with their surroundings, adjusting behaviors and strategies in real time.

\subsection{Physical Computing and Compositionality}
The theory of physical computing, grounded in mathematical formalism, enhances our understanding of designing artificial systems that engage with the complexities of natural intelligence. Insights from category theoretic formalism of physical computing, highlight the adaptability and compositional capabilities essential for artificial general intelligence, emphasizing system compositionality \cite{Dehghani2024}.

This approach mirrors our discussions on the hierarchical and context-sensitive nature of bio-inspired computational models. By focusing on how components compose and interact, we can create systems that fundamentally engage with complex dynamics, akin to biological systems.

\section{Insights from Neuroscience and NeuroAI}
Biological systems can continue to inspire new paradigms in AI, offering principles that could revolutionize our approach to building intelligent systems \cite{Zador2023}.

\subsection{Embracing Embodiment}
While the importance of embodiment in biological intelligence has been acknowledged \cite{Varela2017}, recent neuroscience research provides deeper insights into how physical interaction with the environment shapes the brain as the engine of cognition. Studies show that neural development and synaptic connectivity are profoundly influenced by sensorimotor experiences \cite{Seidler2017,Nwabudike2024, Jenks2021}, underscoring the need for AI systems that can physically interact with their surroundings to achieve adaptive and context-aware intelligence \cite{Steels1995,Pfeifer2007}.



This hints at the embodied Turing test as a more meaningful benchmark for AI. In fact, as discussed earlier, a general problem with the current use of the Turing test is that the goalpost is constantly moved as the definition of the `target' is reinterpreted whenever a new innovative tool, such as neural networks, surpasses the current threshold \cite{Mitchell2024a}.

The embodied Turing test proposes evaluating AI based on its ability to interact physically and adaptively with the world, rather than solely through disembodied tasks like language processing or game playing. This shifts the focus to dynamic, context-dependent interactions that are more representative of natural intelligence. It emphasizes building systems that can sense, act, and adapt to their environments, echoing the evolutionarily honed sensorimotor capabilities of living creatures.

\subsection{Neuromorphic Efficiency and Adaptability}
Although brain comprises roughly $2\%$ of the body weight, it utilizes about $20\%$ of the body's energy intake \cite{Raichle2002}. Despite this significant energy use relative to its size, the brain is remarkably energy-efficient compared to traditional computing models.  Consuming only about 20 watts of power—with approximately 100 billion neurons with $10^15$ synapses firing at $1Hz$ (using $10^{-10}$ Joules per act on potential and $10^{-14}$ Joules per synaptic transmission)—the human brain outperforms supercomputers in tasks like complex pattern recognition and contextual understanding \cite{Sandberg2016, Lennie2003}.

Neuromorphic engineering aims to mimic the brain's efficiency by implementing principles like sparse coding, where only a small subset of neurons is active at any time, and asynchronous communication, where signals are transmitted only when necessary \cite{Indiveri2015}. This contrasts with conventional processors that consume energy continuously, regardless of computational demand \cite{Roy2019}.

Projects like IBM's TrueNorth chip \cite{Merolla2014} and Intel's Loihi \cite{Davies2018} exemplify efforts to build hardware that emulates neural architectures for improved efficiency and performance. By incorporating these insights, AI can strive toward the flexibility and resilience characteristic of biological intelligence. Neuroscience suggests that dynamic interactions within complex neural networks are central to adaptation and robust learning, emphasizing the potential of neuromorphic designs to reduce the immense computational costs of modern AI.

\subsection{Evolutionary Inspiration}
The brain's layered and modular architecture offers a roadmap for scaling AI without sacrificing stability. Evolution incrementally added complexity to existing regulatory intracellular and neural structures \cite{Dehghani2024}, enabling advanced cognitive abilities while maintaining functionality.

In AI development, an incremental approach can mirror this evolutionary path. Beginning with basic sensorimotor systems and gradually adding complexity allows manageable progress toward AGI. Evolutionary algorithms evolve neural network architectures better suited for specific tasks, mimicking natural selection \cite{Stanley2019}. This method has shown promise in developing AI systems that are both efficient and adaptable.



\section{Empirical Evidence of Bio-inspired Success}
The conceptual foundations we've discussed find support in empirical evidence showcasing the success of bio-inspired AI approaches. Below, we examine key case studies that bridge biological inspiration and practical AI applications.

\paragraph{\textbf{Case Study 1: Mimicking the Visual Cortex—Convolutional Neural Networks (CNNs)}}

\textbf{Overview:}
The visual cortex processes visual information hierarchically; early layers detect simple features like edges and contrasts, while deeper layers integrate these into complex patterns and objects \cite{Felleman1991, Dsouza2022, Rolls2023}. This hierarchical organization allows efficient and robust interpretation of visual stimuli, enabling invariant object recognition \cite{Olshausen1996, Booth1998}.

\textbf{Application in AI:}
Convolutional Neural Networks (CNNs), inspired by this biological architecture, have become the backbone of the deep learning revolution \cite{LeCun1998, LeCun2015}. They consist of multiple layers performing convolutions, pooling, and non-linear activations to extract hierarchical features from input images \cite{LeCun1998}. Each layer captures increasingly abstract representations, from raw pixels to high-level concepts.

Yamins et al. \cite{Yamins2014} demonstrated that CNNs modeled after the primate visual cortex could perform object recognition tasks with high accuracy and predict neural responses in the inferior temporal cortex of macaques. By aligning artificial network layers with biological ones, they showed a remarkable correspondence between CNN activations and neural activity.

\textbf{Connection to Conceptual Foundations:}
This case exemplifies the power of \textit{hierarchical organization} and \textit{context-dependent computation} discussed earlier. By mimicking the multi-scale processing of the visual cortex, CNNs harness hierarchical structures for efficient abstraction and pattern recognition. The success of CNNs underscores the importance of embracing biological principles to enhance AI capabilities.

\textbf{Outcomes and Benefits:}

\begin{itemize}
    \item \textbf{Improved Performance:} CNNs have achieved state-of-the-art results in image classification, object detection, and segmentation tasks \cite{Krizhevsky2012,Long2015,LeCun1998}.
    \item \textbf{Biological Plausibility:} Aligning AI models with biological structures enhances our understanding of both artificial and natural intelligence \cite{Yamins2014}.
    \item \textbf{Scalability:} Hierarchical architectures allow CNNs to handle high-dimensional data efficiently \cite{LeCun2015}.
\end{itemize}

\paragraph{\textbf{Case Study 2: Adaptive Robots and Evolving Behaviors—-Xenobots}}
\textbf{Overview:}
Biological organisms exhibit remarkable adaptability, navigating complex environments through flexible behaviors developed over evolutionary timescales. Replicating this adaptability in artificial systems can lead to robots capable of handling unpredictable scenarios.

\textbf{Application in AI:}
Xenobots are novel living robots constructed from frog (\textit{Xenopus laevis}) embryonic cells \cite{Kriegman2020, Kriegman2021}. Using evolutionary algorithms, various configurations were simulated to identify designs capable of locomotion and task performance. The physical xenobots, assembled based on these designs, displayed behaviors such as moving toward targets, self-healing, and cooperative work.

\textbf{Connection to Conceptual Foundations:}
This case illustrates the application of \textit{trial-and-error heuristics} and \textit{evolutionary inspiration} highlighted earlier. By leveraging evolutionary algorithms—a form of guided trial and error—the researchers harnessed modularity and hierarchical organization to discover functional designs. The xenobots' adaptability reflects the importance of embodiment and context-dependent interactions in developing intelligent systems.

\textbf{Outcomes and Benefits:}

\begin{itemize}
    \item \textbf{Adaptability:} Xenobots adjust their behaviors in response to environmental changes \cite{Kriegman2021}.
    \item \textbf{Self-Repair:} They can recover from damage, demonstrating resilience \cite{Kriegman2020, Ball2020}.
    \item \textbf{Applications:} As living robots, they offer potential for intelligent drug delivery, environmental sensing, and understanding morphogenesis \cite{Ball2020, Kriegman2021}.
\end{itemize}

\paragraph{\textbf{Case Study 3: Bridging the Gap—Neuro-inspired Transformers}}
\textbf{Overview:}
Large Language Models (LLMs), such as GPT-4, have revolutionized natural language processing by generating human-like text, translating languages, and answering complex questions \cite{Sejnowski2023,Mitchell2023}. The Transformer architecture underpins these models, utilizing mechanisms like self-attention to process sequential data efficiently \cite{Vaswani2017,Lin2022}.

\textbf{Application in AI:}
Astrocytes, once seen merely as support cells to neurons, are now recognized as key players in cortical computation \cite{Bazargani2016, Kastaneka2020}. Comprising a substantial portion of brain cells (50–90\%), astrocytes interact extensively with numerous synapses and other astrocytes, serving as integrators in spatiotemporal signaling \cite{deCeglia2023}. This connectivity parallels the Transformer architecture, where integration of diverse data inputs optimizes computational efficiency and accuracy \cite{Vaswani2017}.


Recent work suggests parallels between the Transformer architecture and neuron-astrocyte interactions in the brain \cite{Kozachkov2023}. This novel bio-inspired architecture posits that astrocytes implement functions analogous to the self-attention mechanism in Transformers, influencing information flow based on context \cite{Kozachkov2023, Kozachkov2024}.

\textbf{Connection to Conceptual Foundations:}
This case aligns with \textit{context-dependent processing} and \textit{multi-scale interactions} discussed earlier. Recognizing the role of astrocytes acknowledges the importance of non-neuronal components in neural computation, reflecting a holistic view of intelligence. This insight emphasizes the significance of \textit{modularity} and \textit{hierarchical organization} in both biological and artificial systems.

\textbf{Outcomes and Benefits:}
This neuron-astrocyte interaction model opens exciting avenues for building bio-inspired Transformers that could exhibit:
\begin{itemize}
    \item \textbf{Enhanced Learning:} Incorporating astrocyte-like elements could improve adaptive learning in Transformers, allowing more efficient and dynamic learning from new data.
    \item \textbf{Biological Plausibility:} A biologically grounded architecture mirrors accurate neural processes, increasing our understanding of complex cognitive tasks in the brain.
    \item \textbf{Robustness and Generalization:} Modulating synaptic activity through astrocyte-like mechanisms may lead to AI models more resilient to noise and capable of generalizing from limited or varied data.
    \item \textbf{Energy Efficiency:} Bio-inspired designs drawing from astrocyte energy management functions could reduce computational demands, making AI systems more energy-efficient.
\end{itemize}

\paragraph{\textbf{Bridging Bio-inspired AI and Large Language Models}}

\textbf{The Challenge:}
LLMs like GPT-4 spark debates about whether they truly understand language or merely mimic understanding through statistical patterns \cite{Mitchell2023, Sejnowski2023}. While these systems excel at generating coherent text, they often lack deep comprehension, common sense, and the ability to reason causally \cite{Kiciman2024, Cuskley2024, Bender2021}. They may require careful evaluation to prevent deceitful outputs and fallibility \cite{Collins2024, Hagendorgg2024}.

\textbf{Integrating Bio-inspired Concepts:}
Marrying bio-inspired principles with LLMs offers pathways to address these limitations:
\begin{itemize}
    \item \textbf{Hierarchical Structures:} 
    Incorporating biological hierarchies into LLMs could refine contextual and adaptive processing. Training LLMs to build and refine internal concept representations hierarchically, mirroring how brains organize knowledge, may enhance reasoning and abstraction.
    \item \textbf{Experience-Driven Learning:} LLMs thrive on pattern recognition from massive datasets. Bio-inspired approaches emphasize learning through interaction with the environment. Enabling LLMs to engage with simulated environments or multimodal data (e.g., text, vision, action) could ground their understanding, moving beyond pattern recognition to experiential learning.
    \item \textbf{Causal Reasoning:} Integrating mechanisms for causal inference, inspired by biological cognition, could allow LLMs to understand cause-and-effect relationships, improving their problem-solving capabilities and reducing spurious correlations.
    \item \textbf{Collaborative Intelligence:} Combining statistical models with bio-inspired modules may leverage the strengths of both. For example, coupling an LLM with a reinforcement learning agent could enable decision-making that considers vast knowledge and contextual adaptability.
\end{itemize}

\textbf{Connection to Conceptual Foundations:}
This interdisciplinary approach reflects embodied intelligence and evolutionary inspiration. By integrating bio-inspired mechanisms, we address the limitations of purely statistical models, moving toward AI systems that exhibit true general intelligence and contextual understanding.

\textbf{Outcomes and Benefits:}

\begin{itemize}
    \item \textbf{Richer Understanding:} AI models gain deeper comprehension, moving beyond surface-level patterns.
    \item \textbf{Adaptability:} Systems adjust to new tasks and environments more effectively.
    \item \textbf{Alignment:} AI that understands context and causality is better equipped to align with human values and norms.
\end{itemize}

\paragraph{\textbf{Summary}}
These case studies demonstrate the tangible benefits of incorporating bio-inspired principles into AI design. By aligning artificial systems with the hierarchical, context-sensitive, and adaptive nature of biological intelligence, we can overcome limitations inherent in traditional AI paradigms. The empirical successes highlighted here validate the conceptual foundations discussed earlier and illuminate the path forward for developing more robust, efficient, and intelligent AI systems.

\section{Comparative Analysis of Bio-inspired AI Approaches}
Bio-inspired AI distinguishes itself from other AI methodologies—such as symbolic AI, connectionist approaches, evolutionary computation, and hybrid methods—by striving to emulate the comprehensive complexity of biological intelligence. It emphasizes adaptive learning, emergent behaviors, hierarchical information processing, and context-dependent responses, mirroring the adaptability of natural systems. While symbolic AI relies on explicit rules and excels in structured domains, and connectionist approaches focus on pattern recognition using neural networks, bio-inspired AI seeks deeper mimicry, including dynamic reconfiguration and plasticity. Evolutionary computation targets optimization with predefined goals, whereas bio-inspired AI aims for open-ended evolution and self-organization leading to intelligence emergence. Integrating bio-inspired principles with strengths from other methodologies—such as the precision of symbolic AI, the pattern recognition of neural networks, and the optimization of evolutionary algorithms—can enhance adaptability and robustness, accelerating progress toward more intelligent and versatile AI systems.

\section{Methodological Insights from Biology for AI}
Methodological advances in bio-inspired AI draw from biological systems to inform computational models. The design of reconfigurable organisms like Xenobots \cite{Kriegman2020} demonstrates how evolutionary algorithms can navigate complex design spaces, leading to novel configurations that emphasize efficiency, robustness, and adaptability. This highlights evolution as a powerful design tool and underscores the importance of embodiment and physical realization in AI. Our ``HomeoDynamic'' project illustrates how evolutionary optimization enhances reservoir computing models, improving prediction accuracy in chaotic systems through optimized, sparse network structures that reflect biological efficiency. Additionally, hybrid models like the Hopfield-Transformer \cite{Ramsauer2021,Krotov2023} combine associative memory with sequential processing, inspired by hippocampal structures, to enhance memory retrieval and contextual understanding. These examples showcase how biological insights lead to novel computational models, reinforcing the value of interdisciplinary collaboration in advancing bio-inspired AI.

\section{Conclusion}
This paper underscores the crucial role of biological inspiration in developing intelligent artificial systems. Biological intelligence features context-dependent, adaptive behavior emerging from a constantly changing environment. Mimicking this requires a multi-scale organization of information, where causal interactions flow both top-down and bottom-up across different levels, akin to the concept of ``biological relativity'' \cite{Noble2012}. The requisite variety and trial-and-error approaches in nature highlight the fundamental role of experimentation and adaptation—elements often neglected in formal theories of intelligence.

Inspired by Vannevar Bush's influential essay ``As We May Think'' \cite{Bush1996}, we envision a future where computers seamlessly support human endeavors. Achieving this goal depends on a bio-inspired approach that fully embraces the contextual nature of cognition and intelligence. Evolution offers a time-tested blueprint: success through trial and error, adaptability to the environment, and elegant solutions emerging from complexity. The path forward lies in deciphering these principles and translating them into the design of intelligent artificial systems.

\section*{Acknowledgment}
ND wishes to thank Dmitry Krotov, Leo Kozachkov and Aran Nayebi for helpful discussions. ML is grateful for the support of  the Army Research Office via Grant Number W911NF-23-1-0100. The views and conclusions contained in this document are those of the authors and should not be interpreted as representing the official policies, either expressed or implied, of the Army Research Office or the U.S. Government.

\section*{References}
\bibliographystyle{plain}
\bibliography{biblio}

\end{document}